\documentclass[a4paper,11pt]{article}
\usepackage{jcappub,bm,color} 
\usepackage{amssymb,amsfonts,slashed,amsthm,amsmath,graphicx, soul}
\usepackage{comment}
\usepackage{booktabs}
\usepackage{multirow}
\usepackage{makecell}
\usepackage{graphicx}
\usepackage{subcaption}
\usepackage{xcolor}
\usepackage{hyperref}
\usepackage{cleveref}
\usepackage{float}

\makeatletter

\newcommand{\Rmnum}[1]{\expandafter\@slowromancap\romannumeral #1@}
\makeatother

\title{Oscillon decay via parametric resonance: the case of three-point scalar interactions}

\author[a,b]{Siyao Li}

\affiliation[a]{\it Department of Physics, Institute of Science Tokyo,\\Tokyo 152-8551, Japan}
\affiliation[b]{\it Cosmology, Gravity and Astroparticle Physics Group, \\Center for Theoretical Physics of the Universe,\\ Institute for Basic Science, Daejeon 34126, Korea}

\emailAdd{li.s.ap@m.titech.ac.jp}
\date{}

\abstract{
We investigate the decay dynamics of oscillons through interactions with an external scalar field. 
To examine how robust the decay dynamics of oscillons via parametric resonance we previously found in Ref.~\cite{li_decay_2025} are to the specific form of the coupling, we extend the analysis to include a three-point interaction $g_{3}\phi\chi^{2}$. 
We compute the Floquet exponents of the external field $\chi$ under an oscillating oscillon background and analyze how the instability bands depend on the coupling constants and the oscillon shapes. 
Numerical simulations of the two-field system show that, similar to the four-point case, the parametric resonance may cease before the oscillon is destroyed, leaving a smaller oscillon that decays only perturbatively. 
This indicates that the partial decay of oscillons through parametric resonance is a generic phenomenon of oscillon-scalar couplings, qualitatively insensitive to the specific interaction form, while the shape of instability bands, parameter dependence, and the precise critical oscillon energies depend on the specific coupling.
Our findings provide further insights into the decay dynamics of oscillons and their potential role in the post-inflationary reheating process.}

\begin{document}
\maketitle

\section{Introduction}\label{sec:introduction}
Oscillons are localized, oscillating soliton solutions of a real scalar field in a nonlinear potential \cite{lee_nontopological_1992}.
They represent spatially confined oscillatory field configurations that can persist for an exceptionally long time due to the suppression of their radiative decay channels \cite{kasuya_i-balls_2003,gleiser_analytical_2008,fodor_small_2008,fodor_computation_2009,fodor_radiation_2009,hertzberg_quantum_2010,amin_flat-top_2010,salmi_radiation_2012,mukaida_correspondence_2014,saffin_oscillon_2014,mukaida_longevity_2017,ibe_decay_2019,ibe_fragileness_2019,antusch_properties_2019,gleiser_resonant_2019,zhang_classical_2020,olle_recipes_2021,zhang_gravitational_2021, evslin_quantum_2023,wang_excited_2023}.
Their longevity makes them an important nonlinear phenomenon in various cosmological contexts.

During the preheating stage, which is the early, non-perturbative phase of reheating era following inflation \cite{kofman_reheating_1994, shtanov_universe_1995, kofman_towards_1997,garcia-bellido_preheating_1998, micha_turbulent_2004, bassett_inflation_2006}, the inflaton field oscillates coherently about the minimum of its potential \cite{guth_inflationary_1981, linde_new_1982, albrecht_cosmology_1982}.
Parametric resonance during this period can lead to the fragmentation of the inflaton condensate and the formation of oscillons in a wide range of inflationary models \cite{gleiser_pseudostable_1994, amin_inflaton_2010, amin_oscillons_2012, gleiser_generation_2011, lozanov_end_2014, kawasaki_i-ball_2014, amin_nonperturbative_2015,antusch_parametric_2016,lozanov_self-resonance_2018, amin_formation_2019, kou_oscillon_2020,piani_preheating_2023, mahbub_oscillon_2023,drees_inflaton_2025,jia_nonlinear_2025}.
Oscillons have also been observed in other settings, such as cosmological phase transitions \cite{copeland_oscillons_1995, adib_long-lived_2002, farhi_emergence_2008}, axion misalignment and dark matter \cite{kolb_nonlinear_1994,kawasaki_oscillon_2020,arvanitaki_large-misalignment_2020,kawasaki_oscillons_2021,cyncynates_nonperturbative_2022}.

Given that couplings between the inflaton field and other fields are essential for reheating, oscillons are expected to interact with external fields as well.
Such couplings can change the decay rate of oscillons by enabling decay through the particle production of the external fields \cite{farhi_oscillon_2005, graham_electroweak_2007, hertzberg_quantum_2010, gleiser_generation_2011, kawasaki_decay_2014,antusch_impact_2016,van_dissel_symmetric_2022,cyncynates_nonperturbative_2022,shafi_formation_2024, piani_ephemeral_2025}.
Especially, parametric resonance may occur due to the oscillating configurations of the oscillons when coupled to bosonic fields and lead to fast decay.
However, unlike the homogeneous inflaton background, the spatial localization of oscillons can suppress the resonance, as the produced particles can escape from the oscillon, preventing Bose enhancement.
Consequently, exponential amplification of the external field occurs only if the coupling or the background oscillation is strong enough for particle production to outpace the escape \cite{hertzberg_quantum_2010, kawasaki_decay_2014}.

In our previous work \cite{li_decay_2025}, we investigated oscillon decay and lifetime in the presence of a four-point coupling term, $g_4\phi^2\chi^2$. 
We analyzed how the instability bands of the external field $\chi$ depend on its mass, the coupling strength, and the oscillon profile in details.
Our simulations showed that when the oscillon energy decreases below a critical scale, the resonance can terminate before complete destruction of the oscillon, allowing a residual oscillon to survive.
This result provides a natural explanation for the lattice simulation reported in Ref.~\cite{shafi_formation_2024}.

Building upon these findings, a natural question arises: to what extent do these results depend on the specific form of the interaction?
In particular, if a different coupling scheme is adopted, which features of the oscillon decay dynamics remain robust, and which alters qualitatively?
To address this question, in the present work we extend the analysis to include a three-point interaction term, $g_3 \phi \chi^2$. 
We analyze the dependence of the Floquet exponents of $\chi$ field on the coupling constant $g_3$ and the shape of oscillon profile.
Compared to the four-point case, the structure of the instability bands is simpler because of the independent parameters in the Mathieu's equation.
We also perform dynamical simulations of the coupled oscillon-scalar-field system for both $g_3 \neq 0, g_4 = 0$ and $g_3 \neq0, g_4 \neq 0$.
In agreement with our previous study, we observe similar decay behavior of the oscillon: in some cases the parametric resonance occurs and terminates before destroying the oscillon, which implies the fact that the survival of small oscillons after explosive decay into external scalar field is a universal phnomenon qualitatively independent of specific form of oscillon-scalar coupling.

The rest of the paper is organized as follows.
In section \ref{sec:models}, we introduce the model and normalizations used in numerical computations.
In section \ref{sec:single-field oscillon}, we very briefly revisit the single-field oscillons.
In section \ref{sec:instability bands}, we analyze the instability bands of the external scalar field $\chi$ by solving the homogeneous Mathieu's equation and involving the spatial-dependent oscillon profile by numerical simulations.
In section \ref{sec:two-field simulation}, we perform full simulations of two fields under spherical symmetry.
In section \ref{sec:conclusion}, we summarize and conclude our work.

\section{Field models and normalizations}\label{sec:models}
In this work, we consider a model of two real scalar fields $\phi$ and $\chi$.
The $\phi$ field can be the inflaton field with nonlinear self-coupling, which allows oscillon solutions.
The $\chi$ field is a light spectator field with no nonlinear interaction for simplicity.
The two fields are coupled through $\mathcal{L}_{\textrm{int}} \supset g_3\phi \chi^2 + g_4\phi^2 \chi^2$, which can be used for the reheating process after inflation.
The Lagrangian of the two fields is
\begin{align}
\begin{split}
    \mathcal{L} = &\frac{1}{2}\partial_\mu \phi \partial^\mu \phi -V(\phi) + \frac{1}{2}\partial_\mu \chi \partial^\mu \chi - \mathcal{V}(\chi) - g_3 \phi \chi^2 - g_4 \phi^2 \chi^2,\label{eq:full lagrangian}\\
    V(\phi) &=  \frac{1}{2} m_\phi^2 \phi^2 + V_{\textrm{nl}}(\phi),~~\mathcal{V}(\chi) = \frac{1}{2} m_\chi^2 \chi^2.
\end{split}
\end{align}
Here, $V_{\textrm{nl}}(\phi)$ contains the nonlinear terms of $\phi$. 
We take the vaccum at $V(\phi=0) = 0$.
In the Minkowski spacetime, the equations of motion are derived as
\begin{align}
    \Ddot{\phi} - \nabla^2 \phi + m_\phi^2\phi +\frac{d V_{\textrm{nl}}}{d \phi} + g_3 \chi^2 + 2g_4\phi \chi^2 = 0,\label{eq:full phi eom}\\
    \Ddot{\chi} - \nabla^2 \chi + m_\chi^2 \chi + 2g_3\phi\chi + 2g_4\phi^2\chi= 0.\label{eq:full chi eom}
\end{align}
Here, the dot represents the time derivative.
As an example in this work, we use a sextic polynomial potential for $\phi$,
\begin{align}
     V_{\textrm{nl}}(\phi) = - \lambda \phi^4 + \lambda_6\phi^6,\label{eq:phi6 potential}
\end{align}
where $\lambda, \lambda_6 >0$.

By defining the dimensionless variables:
\begin{align}
    \widetilde{x}^\mu = m_\phi x^\mu, ~ \widetilde{\phi} = \frac{\sqrt{\lambda} \phi}{m_\phi},~\widetilde{\lambda_6} = \frac{m_\phi^2 \lambda_6}{\lambda^2}, ~\widetilde{\chi} = \frac{\sqrt{\lambda} \chi}{m_\phi}, ~\widetilde{m}_\chi = \frac{m_\chi}{m_\phi}, ~\widetilde{g_3} = \frac{g_3}{m_\phi\sqrt{\lambda}},~\widetilde{g_4} = \frac{g_4}{\lambda},\label{eq:dimensionless quantities}
\end{align}
the action can be rewritten as,
\begin{align}
    \mathcal{S} = \frac{1}{\lambda}\int d^4\widetilde{x} &\left( \frac{1}{2} \partial_{\widetilde{\mu}} \widetilde{\phi} \partial^{\widetilde{\mu}} \widetilde{\phi} + \frac{1}{2} \partial_\mu \widetilde{\chi} \partial^\mu \widetilde{\chi} - V(\widetilde{\phi}) - \mathcal{V}(\widetilde{\chi}) - \widetilde{g_3} \widetilde{\phi} \widetilde{\chi}^2  - \widetilde{g_4} \widetilde{\phi}^2 \widetilde{\chi}^2\right),\\
    V(\widetilde{\phi}) &= \frac{1}{2} \widetilde{\phi}^2 - \widetilde{\phi}^4 + \widetilde{\lambda_6} \widetilde{\phi}^6,~~\mathcal{V}(\widetilde{\chi}) =  \frac{1}{2} \widetilde{m}_\chi^2 \widetilde{\chi}^2.
\end{align}
These dimensionless quantities are used in all the numerical computation in this work.

\section{Oscillons in single scalar field model}\label{sec:single-field oscillon}
In this work, we assume that the $\chi$ field becomes important to the dynamics of $\phi$ only after the oscillons consisting of $\phi$ are formed. 
This requires the self-coupling of $\phi$ is stronger than the external coupling.
In this section, we briefly review the properties of single-field oscillons when $g_3=0, g_4=0$.

Under the spherical symmetry,  the equation of motion of $\phi$ is
\begin{align}
    \Ddot{\phi} - \frac{\partial^2 \phi}{\partial r^2} - \frac{2}{r}\frac{\partial \phi}{\partial r} +m_\phi^2 \phi +\frac{d V_{\textrm{nl}}}{d \phi} = 0.\label{eq:spherical eom of phi}
\end{align}
Oscillons are localized, oscillating solutions of $\phi$, which can be decomposed as
\begin{align}
    \phi(t,r) \simeq \phi_{osc} (t,r) + \xi (t,r),~~~\phi_{osc} (t,r) \approx 2\psi(r) \cos{(\omega t)},\label{eq:single frequency ansatz}
\end{align}
where $\psi(r)$ is a localized profile, $\xi$ is the perturbation around the oscillon configuration $\phi_{osc}$ and contains radiation modes oscillating rapidly.
$\phi_{osc}$ can be seen as a projection on the real axis of the localized soliton solution in a $U(1)$ invariant complex scalar field theory under non-relativistic limit, which is associated with a particle number conservation~\cite{mukaida_correspondence_2014, mukaida_longevity_2017,blaschke_oscillons_2025,blaschke_oscillons_2025-1}.
Since the relativistic modes in $\xi$ are tiny but nonzero, the conservation law is not exact in the real field thoery and the oscillon decays slowly through $\xi(t,r)$.
We note that, as a result, the profile $\psi(r)$ and frequency $\omega$ should also be time-dependent, but only on a timescale much longer than the oscillation period.

When solving an oscillon configuration in a short timescale comparable to the period, we can neglect the slow time-dependence of $\psi$ and $\omega$.
Substituting the ansatz in Eq.~\eqref{eq:single frequency ansatz} into Eq.~\eqref{eq:spherical eom of phi}, we can derive the equation for the oscillon profile $\psi(r)$ by multiplying a $\cos(\omega t)$ and taking the time-average over one period,
\begin{align}
    \frac{d^2\psi}{dr^2} + \frac{2}{r} \frac{d\psi}{dr} - \left[ ( m_\phi ^2-\omega^2) \psi +\frac{1}{2}\frac{\partial V_{\textrm{eff}}(\psi)}{\partial \psi}\right] = 0,\label{eq:eom oscillon profile}\\
    V_{\textrm{eff}}(\psi) = \overline{V_{\textrm{nl}}(\phi)} = - 6\lambda \psi^4 + 20\lambda_6\psi^6, \label{eq:Veff phi6}
\end{align}
With the following boundary conditions a solution of oscillon profile should satisfy,
\begin{equation}
    \frac{d\psi(r)}{dr} \bigg|_{r=0} =  0,~~
    \psi(r\to \infty) \to 0,\label{eq:oscillon bc}
\end{equation}
we can obtain $\psi(r)$ for any given value of $\omega$ by solving the boundary value problem.
The trivial configuration $\psi=0$ is always a solution to this equation, however, the desired oscillon profile emerges only if the frequency $\omega$ and the shape of the potential satisfy specific conditions.
The necessary conditions on $\omega$ for the existence of such a localized solution of $\psi(r)$ can be derived as \cite{coleman_q-balls_1985,multamaki_analytical_2000,kasuya_i-balls_2003, mukaida_longevity_2017}
\begin{align}
    \min \left[\frac{2V(\psi)}{\psi^2} \right] < \omega^2 < m_\phi^2.\label{eq:localized potential condition}
\end{align}
This condition also requires that the potential $V(\phi)$ is flatter than the quadratic potential at somewhere away from $\phi=0$.

Since the variation of $\psi(r)$ is an adiabatic process when $ |\psi| \gg |\partial_t\psi|/\omega \gg |\partial_t^2\psi|/\omega^2$, the charge of an oscillon as an adiabatic invariant can be defined as \cite{kasuya_i-balls_2003, kawasaki_adiabatic_2015}
\begin{align}
    I \equiv \frac{1}{\omega}\int d^3x \overline{\dot{\phi}^2} = 8\pi\omega \int dr ~r^2 \psi^2(r),\label{eq:adiabatic charge}
\end{align}
where the overline denotes the time-average over one period of oscillation, and the second equality uses the antasz in Eq.~\eqref{eq:single frequency ansatz}.

The energy of an oscillon averaged over a period can also be computed from the profile by
\begin{align}
    \overline{E} &= \int d^3x \left[ \frac{1}{2} \overline{\dot{\phi^2}} + \frac{1}{2} \overline{(\nabla\phi)^2} + \overline{V(\phi)} \right] \nonumber\\
    &=4\pi \int dr~r^2 \left[\omega^2 \psi^2 + \left( \frac{\partial\psi}{\partial r} \right)^2 + m_\phi^2 \psi^2 + V_{\textrm{eff}}(\psi) \right],
    ~~V_{\textrm{eff}}(\psi) \equiv \overline{V_{\textrm{nl}}(\phi)}.\label{eq:time-average energy from oscillon profile}
\end{align}

\begin{figure}[t]
    \centering        \includegraphics[width=\textwidth]{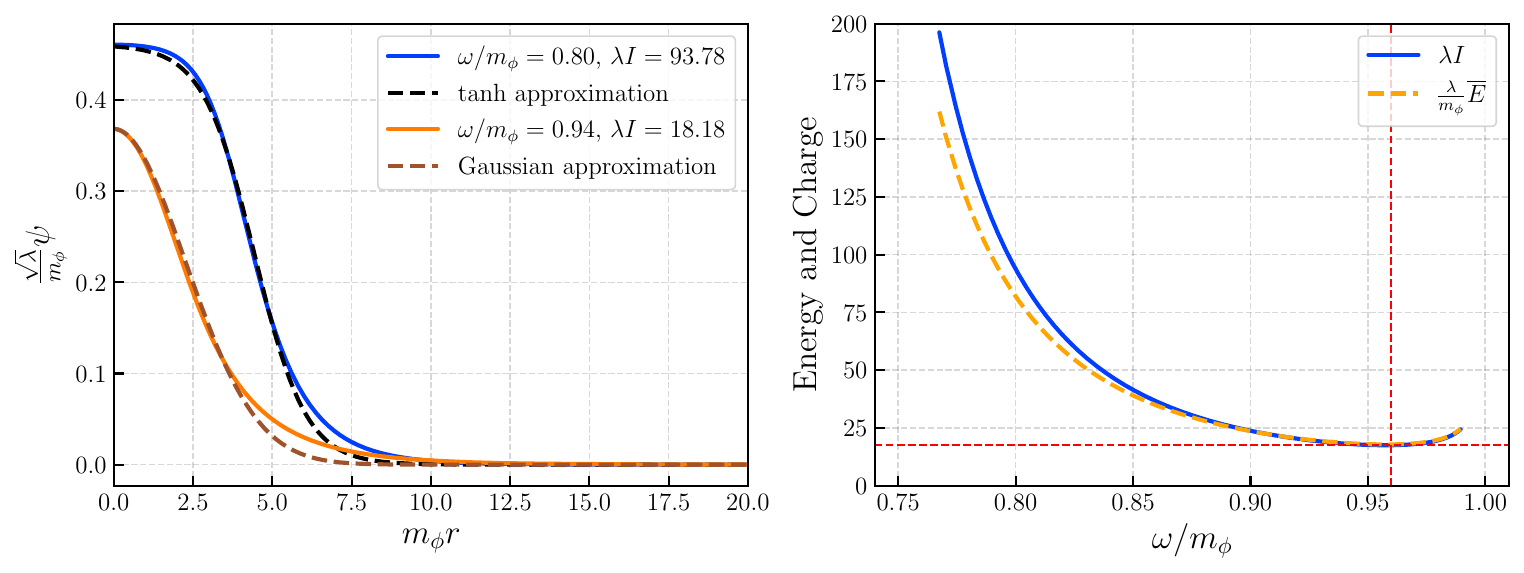}
    \caption{Left panel: Oscillon profiles solved numerically for $\omega/m_\phi= 0.80$ and $\omega/m_\phi= 0.94$, with $m_\phi^2 \lambda_6/\lambda^2 = 0.8$. 
    The corresponding charges $I$ are computed by the integral of profile as given by Eq.~\eqref{eq:adiabatic charge}. 
    Right panel: The dependence of the oscillon charge and energy on its fundamental frequency $\omega$, obtained by solving Eq.~\eqref{eq:eom oscillon profile} for different values of $\omega$ with $m_\phi^2 \lambda_6/\lambda^2 = 0.8$, and integrating Eq.~\eqref{eq:adiabatic charge} and Eq.~\eqref{eq:time-average energy from oscillon profile}.
    The red dashed lines are the critical values of $\omega_{\text{death}}/m_\phi = 0.96$ and $\lambda I_{\text{death}} = 17.48$ for ``energetic death'', beyond which the oscillon solution is no more stable against perturbations. 
    We define the end of the oscillon lifetime as the moment when oscillon reaches this critical value.}
    \label{fig:oscillon profile}
\end{figure}

With the normalization given in Sec.~\ref{sec:models}, the free parameter in the $\phi$ sector is only $m_\phi^2 \lambda_6/\lambda^2$.
Figure~\ref{fig:oscillon profile} shows examples of oscillon profiles solved with $m_\phi^2 \lambda_6/\lambda^2 = 0.8$.
With these profile, we can compute the corresponding charge and energy of the oscillon by integrating Eq.~\eqref{eq:adiabatic charge} and Eq.~\eqref{eq:time-average energy from oscillon profile}.
The relation of oscillon charge and energy with the fundamental frequency $\omega$ is plotted in the right panel of Fig.~\ref{fig:oscillon profile}.
The stability condition in three dimensions gives the upper limit for $\omega$ of a stable oscillon against small perturbations \cite{vakhitov_stationary_1973, lee_nontopological_1992, multamaki_analytical_2000},
\begin{align}
    \frac{d\overline{E}(\omega)}{d\omega} \bigg|_{\omega_{\textrm{death}}} = 0,
\end{align}
Therefore, we use $\omega$ as a parameter which uniquely corresponds to a stable oscillon in this work (we don't consider the `excited' oscillons investigated in Ref~\cite{van_dissel_oscillon_2023}).
Then during the decay process, when the oscillon energy (frequency) reaches the critical value $\overline{E}_{\textrm{death}}$ ($\omega_{\textrm{death}}$), the oscillon experiences ``energetic death'' \cite{cyncynates_structure_2021} and rapidly breaks down into dissipative waves.

Due to the appearence of perturbation $\xi$, oscillons decay simutaneously without external perturbations.
Their decay rate can be derived semi-analytically in terms of profile $\psi(r)$ \cite{mukaida_longevity_2017, ibe_decay_2019,zhang_classical_2020}.
The solution of $\xi$ contains outgoing spherical waves with frequency of integer times of $\omega$, whose amplitudes depends on $\omega$ and $\psi(r)$.
Then the energy decay rate of the oscillon through $\xi$ is computed as 
\begin{align}
    \Gamma_\xi \equiv \frac{1}{\overline{E}} \Bigg|\overline{\frac{dE}{dt}}\Bigg|= 4\pi r^2 \frac{|\overline{\partial_0 \xi \partial_r \xi}|}{\overline{E}}.\label{eq:gamma of single field}
\end{align}

\section{Parametric resonance on oscillon background when  \texorpdfstring{$g_3 \neq 0,\ g_4 = 0$}{g3 != 0, g4 = 0}}\label{sec:instability bands}

The decay and lifetime of oscillons can be affected by the coupling between the oscillon field and other fields \cite{farhi_oscillon_2005, graham_electroweak_2007, hertzberg_quantum_2010, gleiser_generation_2011, kawasaki_decay_2014,van_dissel_symmetric_2022,cyncynates_nonperturbative_2022,shafi_formation_2024, piani_ephemeral_2025}.
In our previous work \cite{li_decay_2025}, we report a detailed analysis of the instability bands of $\chi$ with $g_3 = 0, g_4 \neq 0$.
In this section, we investigate the instability bands in the case of $g_3 \neq 0, g_4 = 0$.
The equations of motion of the two fields are now
\begin{align}
    \Ddot{\phi} - \frac{\partial^2 \phi}{\partial r^2} - \frac{2}{r}\frac{\partial \phi}{\partial r} +m_\phi^2 \phi +\frac{d V_{\textrm{nl}}}{d \phi} + g_3\chi^2 = 0,\label{eq:spherical full phi eom}\\
    \Ddot{\chi} - \frac{\partial^2 \chi}{\partial r^2} - \frac{2}{r}\frac{\partial \chi}{\partial r} + m_\chi^2 \chi + 2 g_3 \phi \chi=0.\label{eq:spherical full chi eom}
\end{align}
Taking the single frequency approximation for oscillon configuration of $\phi$, $\phi(t,r) = 2\psi(r)\cos{(\omega t)}$, the equations of motion for $\chi(t,r)$ becomes
\begin{align}
    \Ddot{\chi} - \frac{\partial^2 \chi}{\partial r^2} - \frac{2}{r}\frac{\partial \chi}{\partial r} + m_\chi^2 \chi+ 4g_3\psi(r) \cos{(\omega t)} \chi = 0.\label{eq:chi eom}
\end{align}
The last term involving the spatial-depenent oscillon profile $\psi(r)$ causes inhomogeneity and mode-mixing of $\chi_k$ with various $k$ values.

\subsection{Homogeneous Floquet analysis}
First, we try to analyze the behaviour of $\chi$ by neglecting the inhomogeneity of $\psi(r)$.
Taking the central amplitude of the oscillon, $\psi(r) \simeq \psi_0 \equiv \psi(r=0)$, the equation can be simplified to
\begin{align}
     \Ddot{\chi_k} + (k^2 + m_\chi^2)\chi_k + 4 g_3 \psi_0 \cos (\omega t) \chi_k = 0, \label{eq:Mathieu's eq with homogeneous}
\end{align}
which can be organized to the form of a standard Mathieu's equation,
\begin{equation}
\begin{aligned}
    &\chi_k^{''} + (A_k + 2 q \cos{(2z)})\chi_k  = 0,\\
    A_k &= \frac{4(k^2 + m_\chi^2)}{\omega^2},~~ q = \frac{8g_3\psi_0}{\omega^2},~~ z = \frac{\omega t}{2},\label{eq:standard Mathieu's eq}
\end{aligned}
\end{equation}
where prime denotes the derivative with respect to $z$.
According to the Floquet's theorem~\cite{floquet_sur_1883,mclachlan_theory_1947}, Mathieu's equation has solutions as
\begin{align}
    \chi_k(z) = \mathcal{P}_{+}(z) e^{\mu z} + \mathcal{P}_{-}(z) e^{- \mu z}, \label{eq:floquet theorem}
\end{align}
where $\mathcal{P}_{\pm}(z)$ are periodic functions with a period of $\pi$ with respect to $z$, and $\mu$ is generally a complex number called the Floquet exponent.
If $\mu$ has a nonzero real part, resonance occurs and part of the solution can get exponentially amplified with time $z$, while the solution is stably oscillating if $\mu$ is purely imaginary.

\begin{figure}[t]
    \begin{subfigure}[t]{0.48\textwidth}
        \centering        \includegraphics[width=\textwidth]{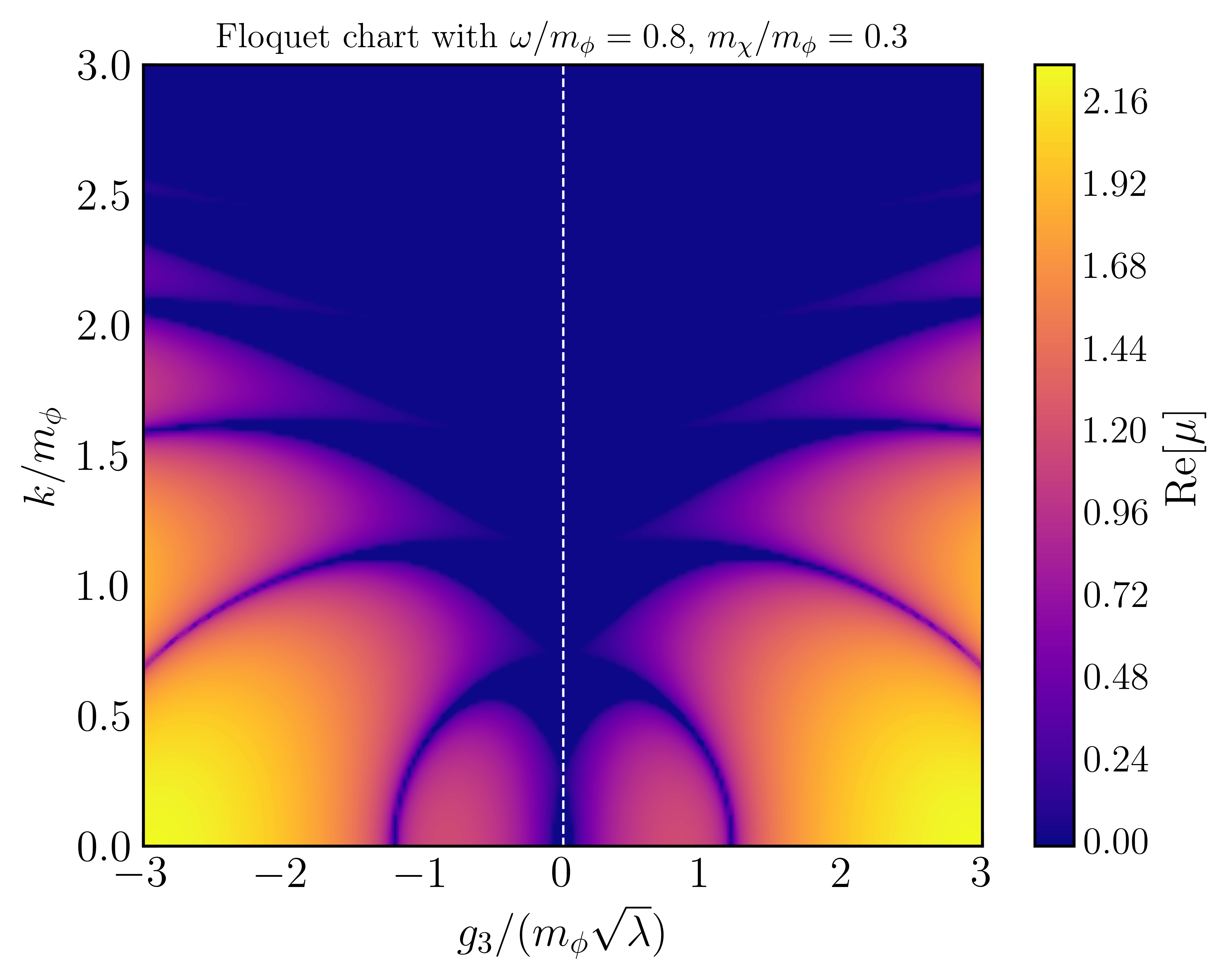}
    \end{subfigure}
    \hfill
    \begin{subfigure}[t]{0.48\textwidth}
        \centering        \includegraphics[width=\textwidth]{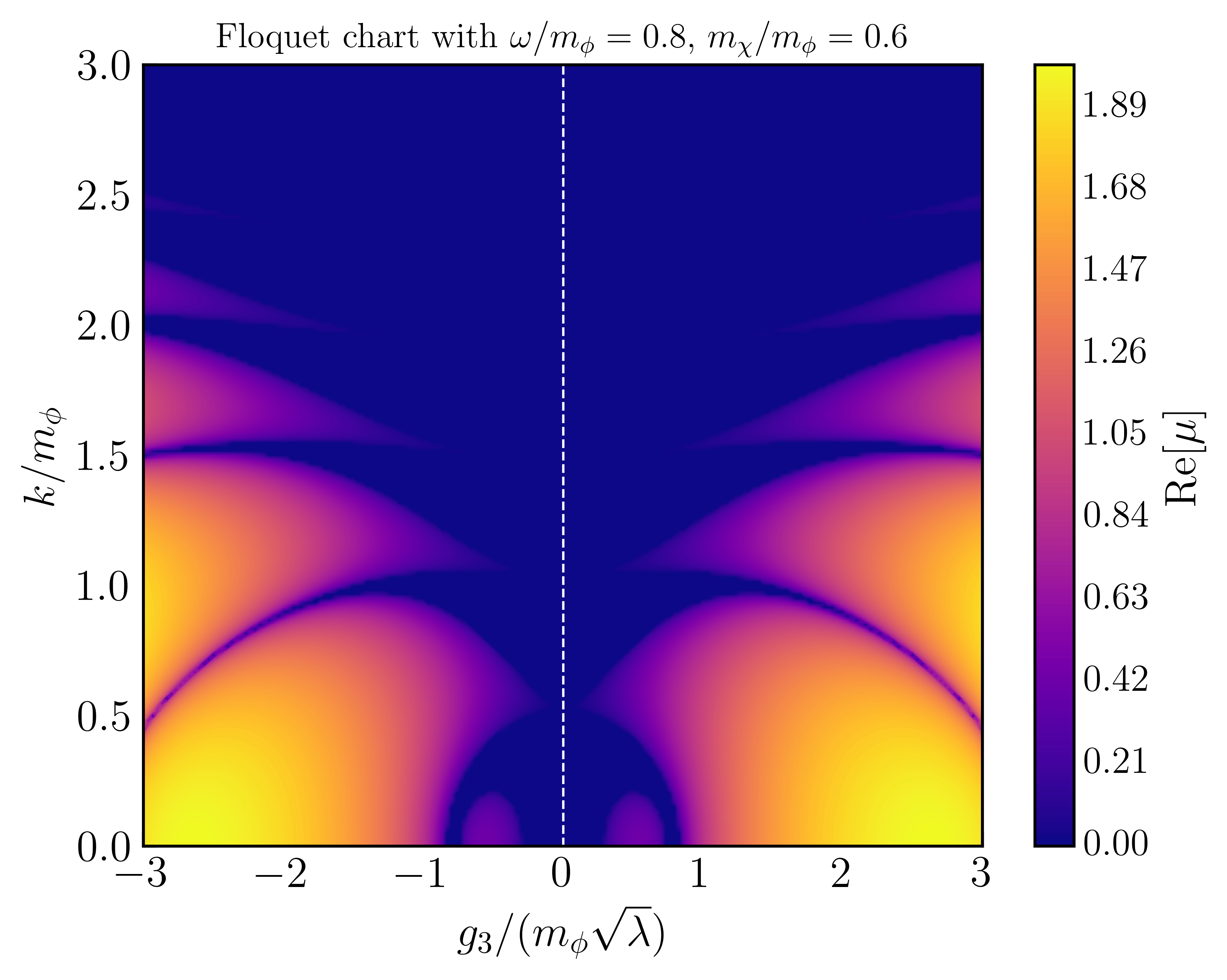}
    \end{subfigure}
    \caption{Floquet charts of the Mathieu equation in Eq.~\eqref{eq:standard Mathieu's eq} neglecting oscillon inhomogeneity, shown in the $(k, g_3)$ plane. 
    The oscillon center amplitude $\psi_0$ corresponds to $\omega/m_\phi = 0.8$. 
    Left and right panels: $m_\chi/m_\phi = 0.3$ and $0.6$, respectively. 
    Dashed line: $g_3 = 0$.}
\label{fig:Floquet_chart_homogeneous}
\end{figure}
\begin{figure}[b]
    \centering        \includegraphics[width=0.5\textwidth]{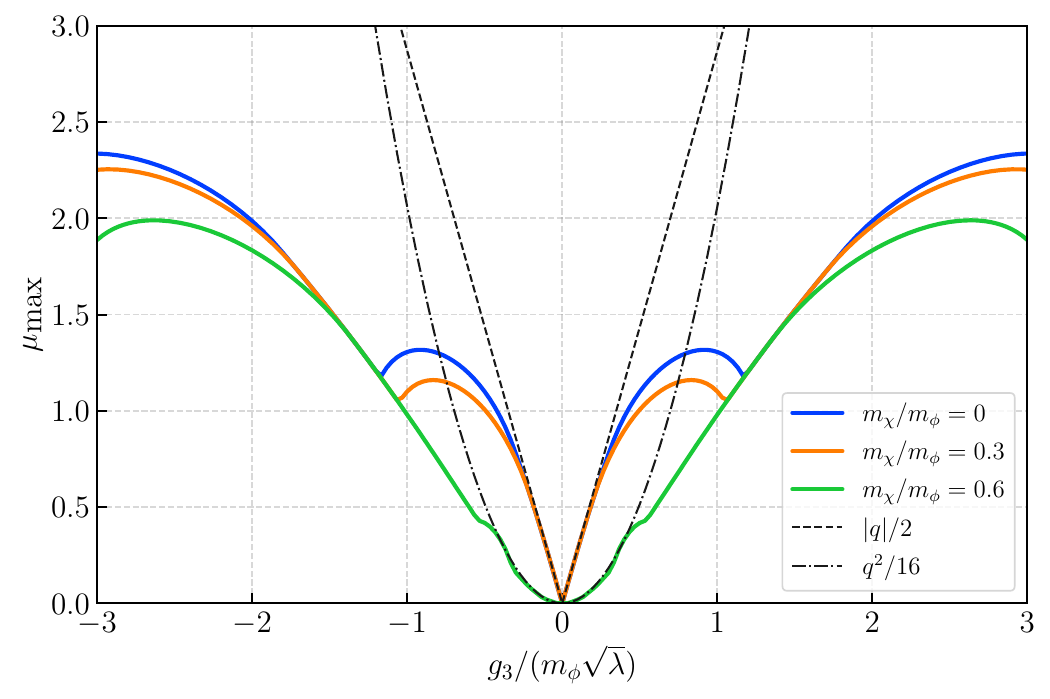}
    \caption{The maximum value of the real part of Floquet exponents, $\mu_{\max} \equiv \max{(\Re{(\mu)})}$, representing the growth rate of the most unstable mode, is shown for various coupling strengths $g_3/(m_\phi\sqrt{\lambda})$ and mass $m_\chi/m_\phi$. 
    The center amplitude $\psi_0 (\simeq 0.46 m/\sqrt{\lambda})$ and frequency of an oscillon with $\omega/m_\phi = 0.8$ are used in the computation. 
    The dashed line denotes the linear dependence in the first narrow resonance band, $\mu_{\max} = |q|/2$, while the dot-dashed line indicates the quadratic dependence in the second narrow band, $\mu_{\max} = q^2/16$.}\label{fig: homogeneous mu max}
\end{figure}

Figure \ref{fig:Floquet_chart_homogeneous} shows the results of Floquet analysis for Eq.~\eqref{eq:standard Mathieu's eq} in terms of various values of $k$ and $g_3$.
Unlike the case of four-point coupling investigated in Ref.~\cite{li_decay_2025}, the parameters $A_k$ and $q$ are independent.
There is always $A_k \geq 0$ so no tachyonic bands appear. 
In addition, since the equation is symmetric about $g_3=0$, behavior of $\chi$ field is the same for attractive and repulsive interaction.

A larger $m_\chi$ makes the Floquet chart shift downwards along the $k$ axis.
When $m_\chi/m_\phi<0.5$, the tree-level decay $\phi \to \chi \chi$ are allowed to occur, which corresponds to the first narrow bands in the limit of $|q|\ll 1$.
As shown in the right panel of Fig.~\ref{fig:Floquet_chart_homogeneous}, the energy conservation forbids the single $\phi$ decay and no mode falls into the first narrow bands when $m_\chi/m_\phi > 0.5$.
However, the $n$-th bands, corresponding to $n$-particle decay, are still allowed when $m_\chi/m_\phi < n/2$.
For instance, the second narrow band is the dominant in the right panel of Fig.~\ref{fig:Floquet_chart_homogeneous} for $m_\chi/m_\phi = 0.6$.  
This can also be seen in Figure~\ref{fig: homogeneous mu max}, where we define $\mu_{\textrm{max}} \equiv \max(\Re{(\mu)})$ as the exponent of the mode that grows fastest on a given background and coupling $g_3$.
From the properties of the Mathieu's equation~\cite{mclachlan_theory_1947}, in the range of small $|g_3|$, we have 
\begin{align}
    \mu_{\textrm{max}} \approx |q|/2 = \frac{4\psi_0}{\omega^2} |g_3|, \quad |g_3| \ll 1\label{eq:narrow band mu relation}
\end{align}
in the first narrow band, while the second narrow band with
\begin{align}
    \mu_{\textrm{max}} \approx q^2/16 = \frac{4\psi_0^2}{\omega^4} g_3^2, \quad |g_3| \ll 1\label{eq:2nd narrow band mu relation}
\end{align}
fits well at small $|g_3|$ when $m_\chi/m_\phi > 0.5$ in Fig.~\ref{fig: homogeneous mu max}.


\subsection{The effect of inhomogeneous oscillon profiles}\label{subsec: bgfix}

Now we investigate the effect of the inhomogeneity of background as an oscillon on the instability bands of $\chi$.

Following our previous work, we perform numerical simulations by evolving $\widetilde{\chi}(t,r)$ on a fixed oscillon background, $\widetilde{\phi}(\widetilde{t},\widetilde{r}) = 2\widetilde{\psi}(\widetilde{r})\cos{(\widetilde{\omega} \widetilde{t})},~\widetilde{\omega} = \omega/m_\phi$ kept fixed at each time step.
Here, $\widetilde{\psi}(\widetilde{r})$ is the numerically obtained profile of a single-field oscillon for various given $\widetilde{\omega}$, as described in Sec.~\ref{sec:single-field oscillon}.
Under spherical symmetry, the three-dimensional dynamics can be effectively reduced to a one-dimensional radial equation.
We numerically solve the nonlinear radial equation of motion given in Eq.~\eqref{eq:spherical eom of phi} within a box of size $r_{\textrm{box}} = 192m_\phi^{-1}$ using 3072 grid points. 
Time evolution is carried out with a  fourth-order symplectic integrator method \cite{yoshida_construction_1990} with time steps $\Delta t = 0.01m_\phi^{-1}$ and spatial derivatives are calculated using a fourth-order central difference.
To eliminate unphysical reflections from the boundaries, we impose an adiabatic damping boundary condition \cite{gleiser_long-lived_2000,gleiser_generation_2011}, and we have confirmed that the results are consistent with those obtained in larger simulation boxes.
All  physical quantities are evaluated within a radius of $r_{\textrm{max}} = 30m_\phi^{-1}$, which is sufficiently large compared to the oscillon radius.

The initial condition for $\widetilde{\chi}$ is specified as a Gaussian profile,
\begin{align} 
    \begin{split} 
        \widetilde{\chi}(0,\widetilde{r}) &= \chi_0 e^{-\widetilde{r}^2/\widetilde{R}_\chi^2},\\ \dot{\widetilde{\chi}}(0,\widetilde{r}) &= 0, 
    \end{split} 
\end{align}
with $\chi_0 = 0.1$ and $\widetilde{R}\chi = 6$.
At the beginning of the simulation, most components of the initial Gaussian profile dissipate rapidly, except for those modes that experience exponential amplification through parametric resonance.
After a short transient stage, the $\chi$ field relaxes into a stable configuration determined predominantly by the fastest-growing mode on the given $\phi$ background.
Therefore, the specific choice of the initial parameters for $\widetilde{\chi}$ does not affect the subsequent growth rate, but only influences the duration of the initial relaxation stage, which is excluded when fitting the exponents.
In this setup, the resonance can drive $\widetilde{\chi}$ to evolve into a localized, oscillon-like configuration due to the nontrivial spatial dependence of the background field $\widetilde{\phi}$.

We then evolve the equation of motion for $\widetilde{\chi}(\widetilde{t}, \widetilde{r})$ in Eq.~\eqref{eq:chi eom}, keeping the oscillon background fixed as
$\widetilde{\phi}(\widetilde{t}, \widetilde{r}) = 2\widetilde{\psi}(\widetilde{r})\cos(\widetilde{\omega}\widetilde{t})$.
The time-averaged energy of the $\chi$ field is computed as
\begin{align} 
\widetilde{\overline{E}}_\chi(\widetilde{t}) = \frac{1}{T_{\textrm{ave}}} \int_{\widetilde{t}}^{\widetilde{t}+T_{\textrm{ave}}} d\widetilde{t} \int_0^{\widetilde{r}_{max}} d\widetilde{r}~ 4\pi \widetilde{r}^2 \left(\frac{1}{2} \dot{\widetilde{\chi}}^2 + \frac{1}{2} (\partial_{\widetilde{r}} \widetilde{\chi})^2 + \frac{1}{2} \widetilde{m}_\chi^2 \widetilde{\chi}^2 \right), \label{eq:chi energy} 
\end{align}
where $T_{\textrm{ave}} = 100m_\phi^{-1}$ is adopted.

Assuming that the fastest-growing mode dominates the energy evolution of $\chi$, the Floquet exponent is extracted by fitting
$\widetilde{\overline{E}}_\chi \propto e^{\Re(\mu)\widetilde{\omega} \widetilde{t}}$,
where $\widetilde{\omega}$ denotes the frequency of the background oscillon.
Since $\widetilde{\overline{E}}_\chi$ is evaluated within a finite domain of radius $\widetilde{r}_{\max}$, if $\chi$ lies in a stable band (i.e., $\Re(\mu)=0$), the energy gradually decreases over time as the initial input propagates beyond the boundary of simulation box through dissipative modes.
We take $\Re(\mu)=0$ whenever $\widetilde{\overline{E}}_\chi$ continues to decrease after the initial relaxation phase in the simulation.

\begin{figure}[t]
    \begin{subfigure}[t]{0.48\textwidth}
        \centering        \includegraphics[width=\textwidth]{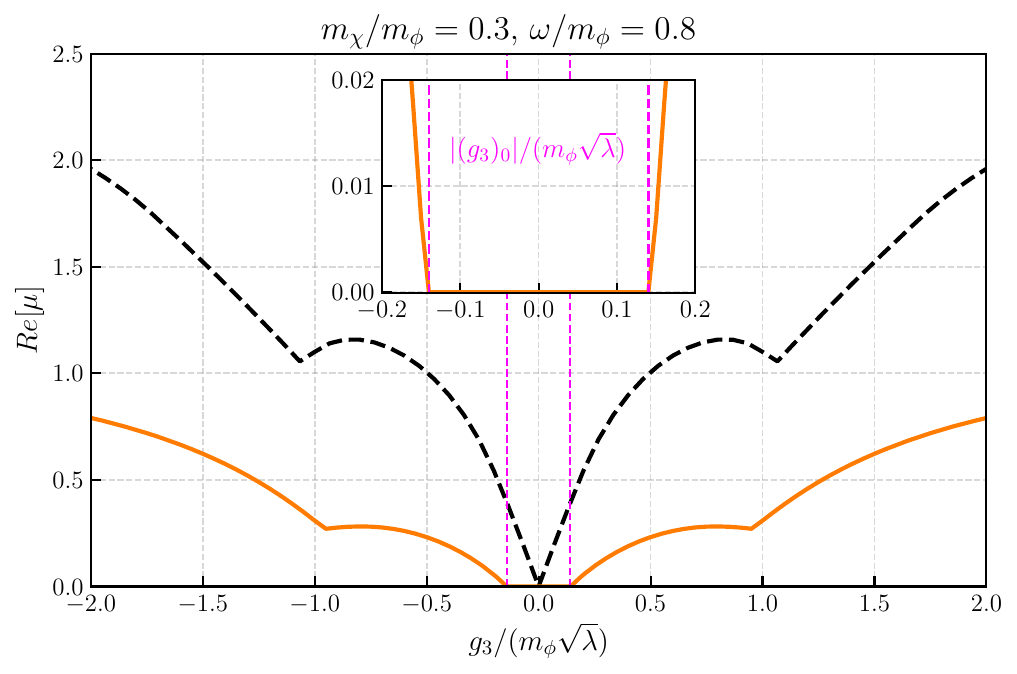}
    \end{subfigure}
    \hfill
    \begin{subfigure}[t]{0.48\textwidth}
        \centering        \includegraphics[width=\textwidth]{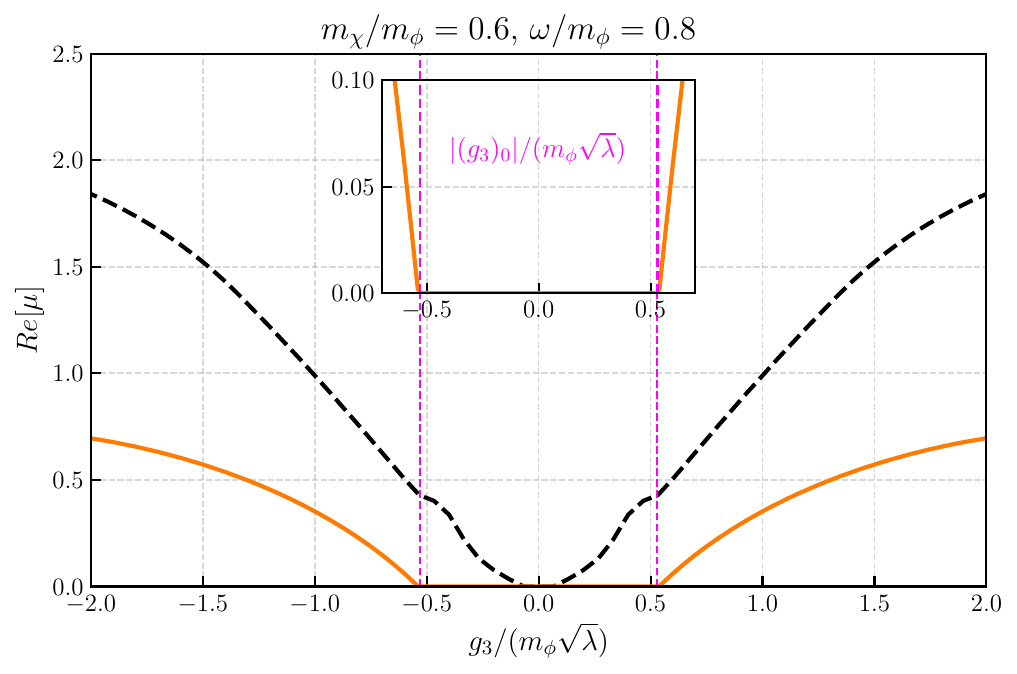}
    \end{subfigure}
    \caption{Growth rate of the $\chi$ field obtained from numerical simulations with a fixed oscillon background of $\omega/m_\phi = 0.8$, for various values of $m_\chi$.
    The black dashed line shows the corresponding $\mu_{\max}$ from the homogeneous Floquet analysis presented in the previous subsection, for comparison. }\label{fig:massive mu compare}
\end{figure}

Figure \ref{fig:massive mu compare} shows the real part of the Floquet exponent, $\Re{(\mu)}$, obtained from numerical simulations with an inhomogeneous oscillon background for various values of $m_\chi$.
As discussed in Refs.~\cite{hertzberg_quantum_2010, kawasaki_decay_2014}, the growth rate of the $\chi$ field is clearly suppressed once the finite-size effect of the oscillon is taken into account compared with the homogeneous case, which can be interpreted as the escape of $\chi$ particles from a localized oscillon of finite radius.
The energy of the produced $\chi$ particles produced can be estimated by
\begin{align}
    E_\chi = nm_\phi/2,\quad n = 1+ \sum_{k=1}^n H \left( \frac{m_\chi}{m_\phi} - \frac{k}{2}\right),
\end{align}
where $H(x)$ denotes the Heaviside step function.
The corresponding escape rate can then be approximated by
\begin{align}
    \Gamma_{\text{escape}} \sim v_\chi/R \sim \frac{p_\chi}{E_\chi R} \sim \frac{1}{R}\sqrt{1-\frac{4}{n^2} \left(\frac{m_\chi^2}{m_\phi^2} \right)},
\end{align}
where we have used $p_\chi \simeq \sqrt{E_\chi^2 - m_{\chi}^2}$, and $R$ is the radius of the background oscillon.

Then we can define a critical value of $|g_3| \sim (g_3)_0$ as the weakest coupling strength for which $\Re{(\mu)} \neq 0$.
The value of $(g_3)_0$ can be estimated from the condition $\Re{(\mu)} \simeq \Gamma_{\text{escape}}$.
When $m_\chi/m_\phi\lesssim 0.5$, this relation gives
\begin{align}
  \frac{2\psi_0}{\omega}|g_3| \simeq  \frac{\sqrt{m^2 - 4m_\chi^2}}{m_\phi R}, \quad m_\chi \lesssim 0.5 m_\phi,
\end{align}
and thus,
\begin{align}
    (g_3)_0 \sim \frac{\omega}{2 \psi_0 R} \sqrt{1- 4 \left(\frac{m_\chi}{m_\phi} \right)^2}, \quad m_\chi \lesssim 0.5 m_\phi. \label{eq:1st g0}
\end{align}
When $m_\chi/m_\phi >0.5$, the right panel of Fig.~\ref{fig:massive mu compare} shows that $(g_3)_0$ exceeds the perturbative regime where Eq.~\eqref{eq:2nd narrow band mu relation} is valid, and therefore the value of $\Re{(\mu)}$ in the second broad resonance band is needed to estimate $(g_3)_0$.

\begin{figure}[t]
    \centering        \includegraphics[width=0.4\textwidth]{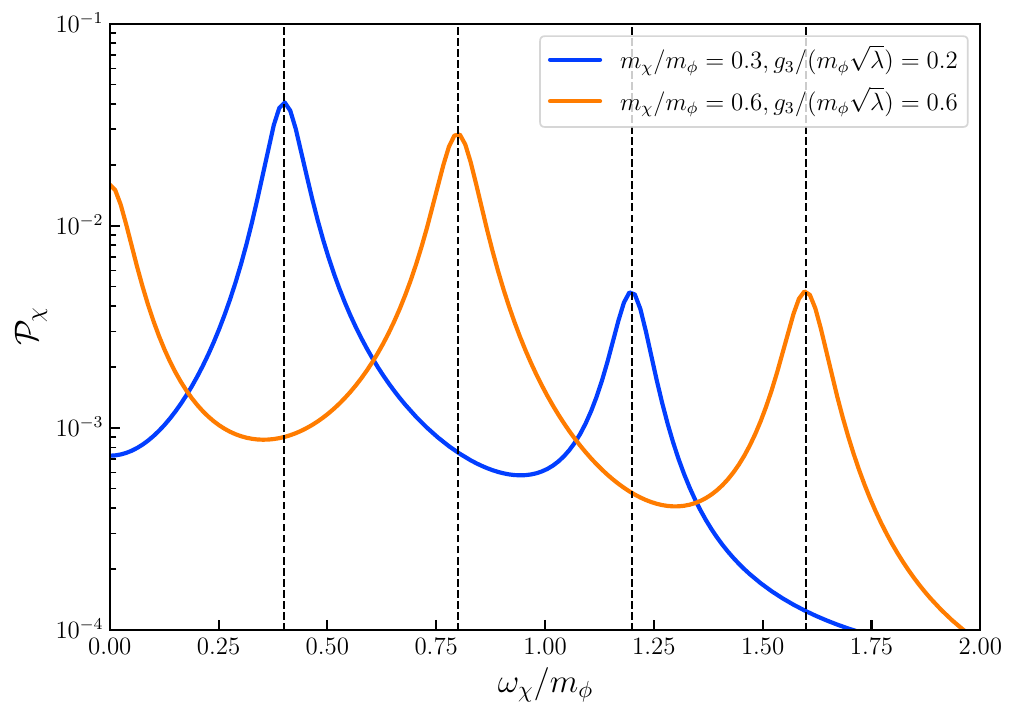}
    \caption{Normalized Fourier spectrum of the center value of $\widetilde{\chi}$, $\widetilde{\chi}(\widetilde{t},0)$, obtained from simulations with a fixed oscillon background $\widetilde{\phi}(\widetilde{t},\widetilde{r})$ for $\omega/m_\phi = 0.8$. Dashed lines indicate the half-integer multiples of the background frequency, $n\omega/2$.}\label{fig:chi fft}
\end{figure}

Figure \ref{fig:chi fft} shows the normalized Fourier spectra of $\widetilde{\chi}(\widetilde{t},0)$ obtained from the simulation.
The dominant peaks appear at $\omega_k = \sqrt{k^2 + m_\chi^2} = n \omega/2$, which correspond to the primary $n$-th narrow parametric resonance occuring at $A_k \simeq n^2$.
The secondary peaks at $(n\pm1)\omega$ arise from the frequency modulation induced by the coupling term.

\begin{figure}[t]
     \begin{subfigure}[t]{0.49\textwidth}
        \centering        \includegraphics[width=\textwidth]{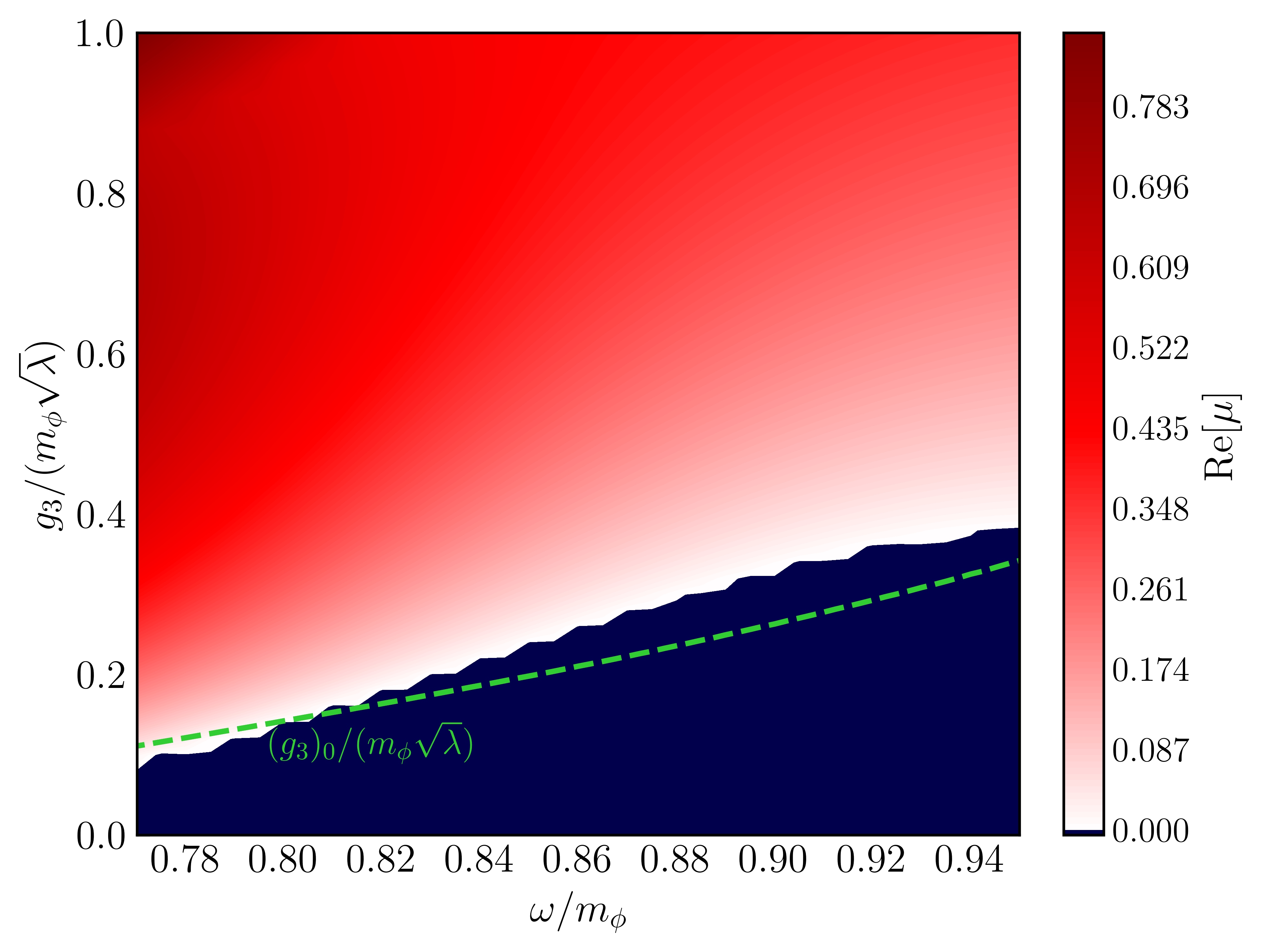}
    \end{subfigure}
    \hfill
    \begin{subfigure}[t]{0.49\textwidth}
        \centering        \includegraphics[width=\textwidth]{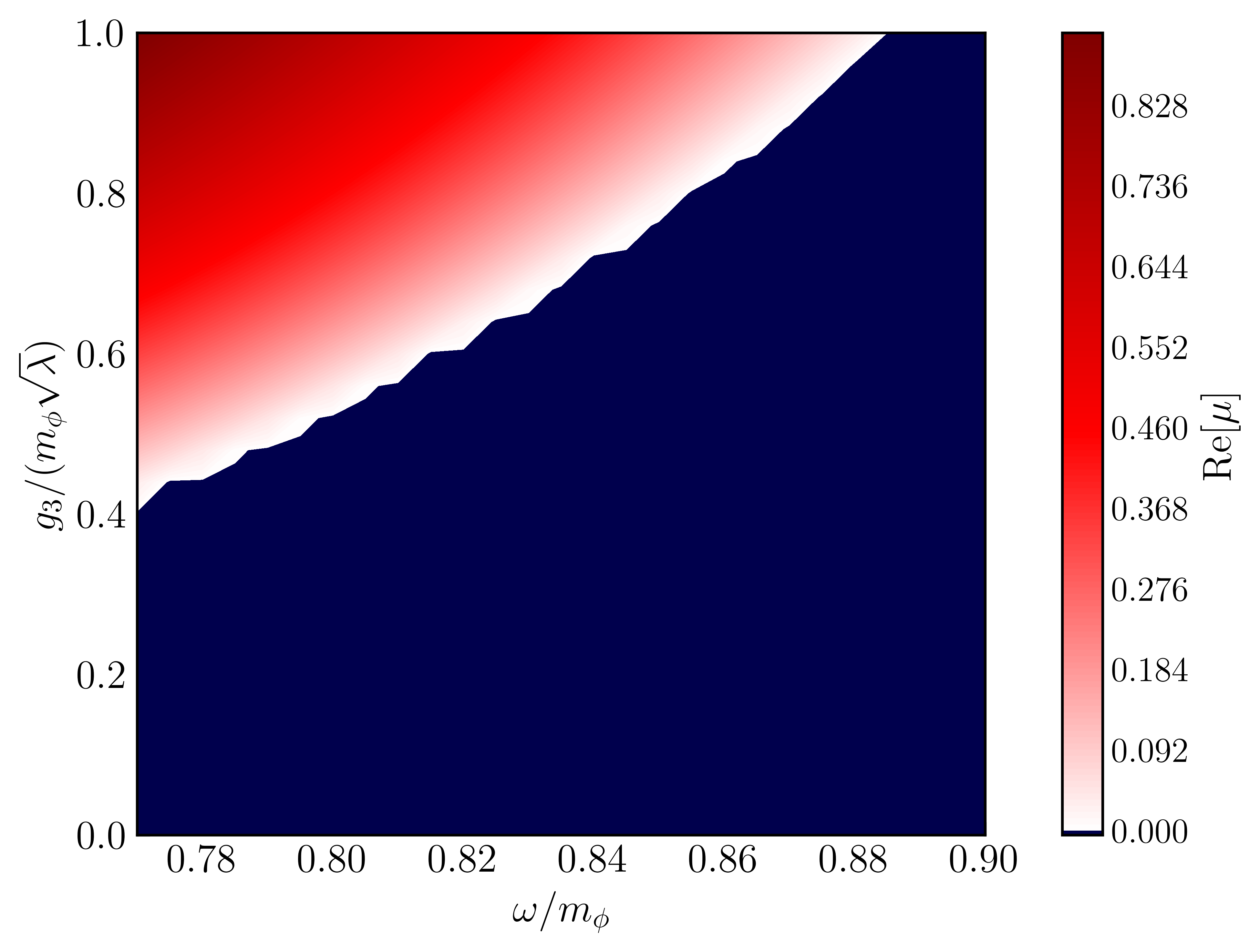}
    \end{subfigure}
    \caption{Contour plots of the exponential growth rate $\Re(\mu)$ of $\chi$ on an oscillon background with for various values of $\omega$, $m_\phi^2 \lambda_6/\lambda^2 = 0.8$ with $m_\chi/m_\phi = 0.3$ (left) and $m)\chi/m_\phi = 0.6$ (right).
    The blue region is the stable region where no visible growth of $\chi$ occurs because of the particle escape. 
    The red region corresponds to the instability bands of parametric resonance.
    The first band corresponds to $\phi \to \chi\chi$ is the main channel in the left panel, while the second band $\phi \phi \to \chi\chi$ is dominant in the right panel.
    The green dashed line plots $(g_3)_0$ given in Eq.~\eqref{eq:1st g0}.}\label{fig:contour mu}
\end{figure}

Finally, we present the exponential growth rate $\Re{(\mu)}$ of $\chi$ obtained from simulations with various oscillon configurations as fixed backgrounds, for $m_\chi/m_\phi = 0.3$ and $m_\chi/m_\phi = 0.6$, in Figure~\ref{fig:contour mu}.
The red regions indicate the onset of parametric resonance for sufficiently strong coupling, $|g_3| > (g_3)_0$.
The green dashed line in the left panel represents the relation of $(g_3)_0$ and the oscillon profile shape given in Eq.~\eqref{eq:1st g0} for $m_\chi/m_\phi =0.3$. 

Simliar to the four-point coupling case we previously studied in Ref.~\cite{li_decay_2025}, a critical value of oscillon energy $\overline{E}_{0}^{\textrm{osc}}$ (corresponding to a critical frequency $\omega_0$) can be defined for a given $g_3$ and $m_\chi$ as
\begin{align}
    \Re(\tilde{\mu}) \begin{cases}
        >0, \quad \overline{E} > \overline{E}_{0}^{\textrm{osc}} ~(\omega<\omega_0)\\
         = 0, \quad \overline{E} \leq \overline{E}_{0}^{\textrm{osc}} ~(\omega\geq\omega_0)
    \end{cases}
\end{align}
where $\overline{E}$ denotes the  oscillon energy obtained from Eq.~\eqref{eq:time-average energy from oscillon profile}.
This implies that during oscillon decay, the resonance of $\chi$ may cease when the oscillon becomes small enough, which is confirmed by our simulation in the next section.

\section{Two-field simulation of an oscillon with external coupling}\label{sec:two-field simulation}
In this section, we perform full numerical simulations for both the time evolution of the $\phi$ and $\chi$ fields with the same numerical setup as in the previous section.
By solving the coupled radial equations of motion for $\phi$ and $\chi$ given in Eq.~\eqref{eq:full phi eom} \eqref{eq:full chi eom}, we investigate the complete decay process of the oscillon in the presence of the external $\chi$ field.
This includes the backreaction of the exponentially growing $\chi$ field on the oscillon, and the growth rate of $\chi$ field on the background of an evolving $\phi$ oscillon.
The initial conditions are specified as
\begin{align}
    \widetilde{\phi}(0, \widetilde{r}) &= 2\widetilde{\psi}(\widetilde{r}),\\
    \widetilde{\chi} (0, \widetilde{r}) &= \widetilde{\chi}_0 \bar{\widetilde{\chi}}(\widetilde{r}),\\
    \dot{\widetilde{\phi}} (0, \widetilde{r})&= \dot{\widetilde{\chi}}(0, \widetilde{r}) = 0,
\end{align}
where $\widetilde{\psi}(\widetilde{r})$ denotes the single-field oscillon profile obtained numerically in Sec.~\ref{sec:single-field oscillon} for a given value of $\widetilde{\omega}$, and $\bar{\widetilde{\chi}}(\widetilde{r})$ is a Gaussian function with a centeral amplitude $\widetilde{\chi}_0 = 0.01$ and radius $\widetilde{R}_\chi = 6$.
Adiabatic damping boundary conditions are imposed on both fields at the outer edge of the simulation box.

The time-averaged energies are computed as 
\begin{align}
    \overline{E}_\phi(\widetilde{t}) &= \frac{1}{T_{\textrm{ave}}} \int_{\widetilde{t}}^{\widetilde{t} + T_{\textrm{ave}}} d\widetilde{t} \int_0^{\widetilde{r}_{max}} d\widetilde{r} ~ 4\pi \widetilde{r}^2 \left(\frac{1}{2} \dot{\widetilde{\phi}}^2 + \frac{1}{2} (\partial_{\widetilde{r}}\widetilde{\phi})^2 + V(\widetilde{\phi}) \right),\\
    \widetilde{\overline{E}}_{\textrm{tot}} (\widetilde{t}) &= \widetilde{\overline{E}}_{\phi} (\widetilde{t})+ \widetilde{\overline{E}}_{\chi}(\widetilde{t}) +\frac{1}{T_{\textrm{ave}}} \int_{\widetilde{t}}^{\widetilde{t}+T_{\textrm{ave}}} d\widetilde{t} \int_0^{\widetilde{r}_{max}} d\widetilde{r}~ 4\pi \widetilde{r}^2 \left(\widetilde{g_3} \widetilde{\phi} \widetilde{\chi}^2 + \widetilde{g_4} \widetilde{\phi}^2\widetilde{\chi}^2\right),
\end{align}
where $\widetilde{\overline{E}}_{\chi}(\widetilde{t})$ is given in Eq.~\eqref{eq:chi energy}.
We take $T_{\textrm{ave}} = 20 m_\phi^{-1}$ and $\widetilde{r}_{max} = 30$.
Note that the total energy should remain constant at least until $\widetilde{t} = \widetilde{r}_{max}$, after which it may begin to decrease due to dissipative modes propagating beyond the monitored region of size $\widetilde{r}_{max}$.

\begin{figure}[b]
     \begin{subfigure}{0.49\textwidth}
        \centering        \includegraphics[width=\textwidth]{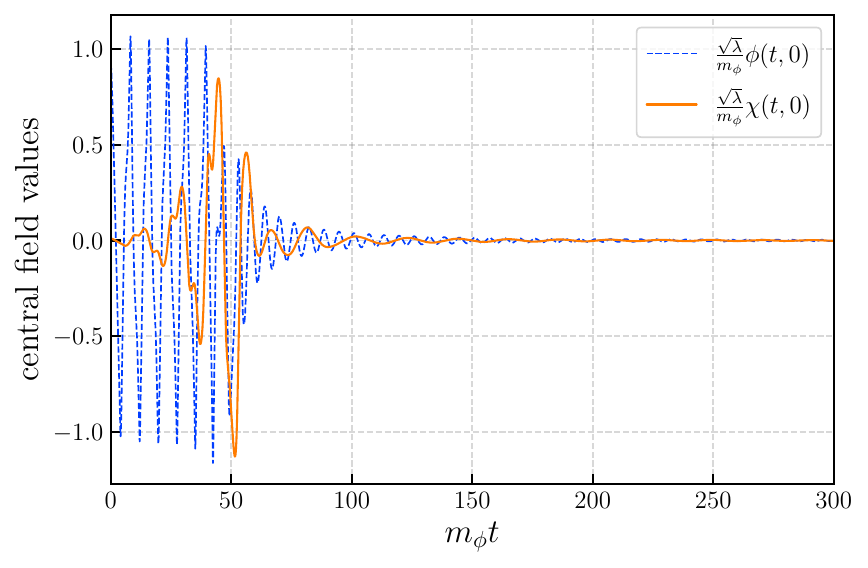}
    \end{subfigure}
    \hfill
    \begin{subfigure}{0.49\textwidth}
        \centering        \includegraphics[width=\textwidth]{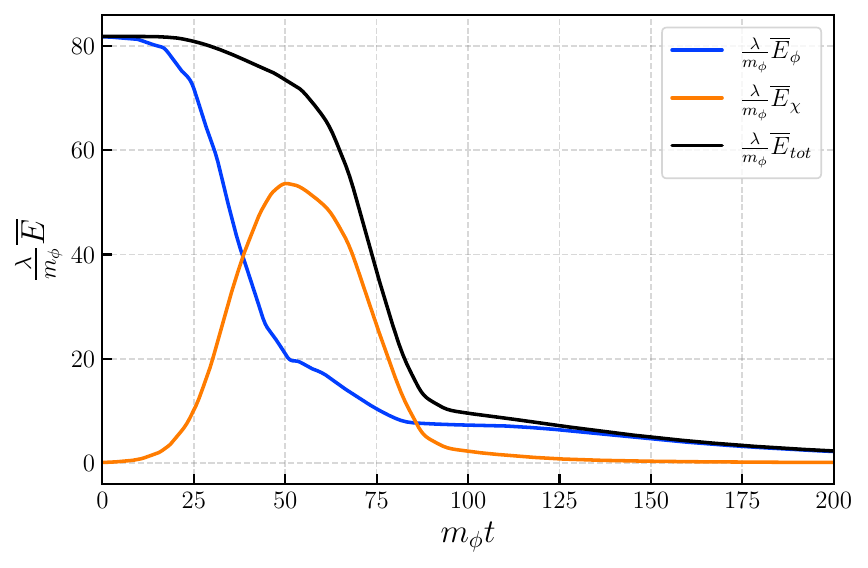}
    \end{subfigure}
    \vfill
    \begin{subfigure}{0.49\textwidth}
        \centering        \includegraphics[width=\textwidth]{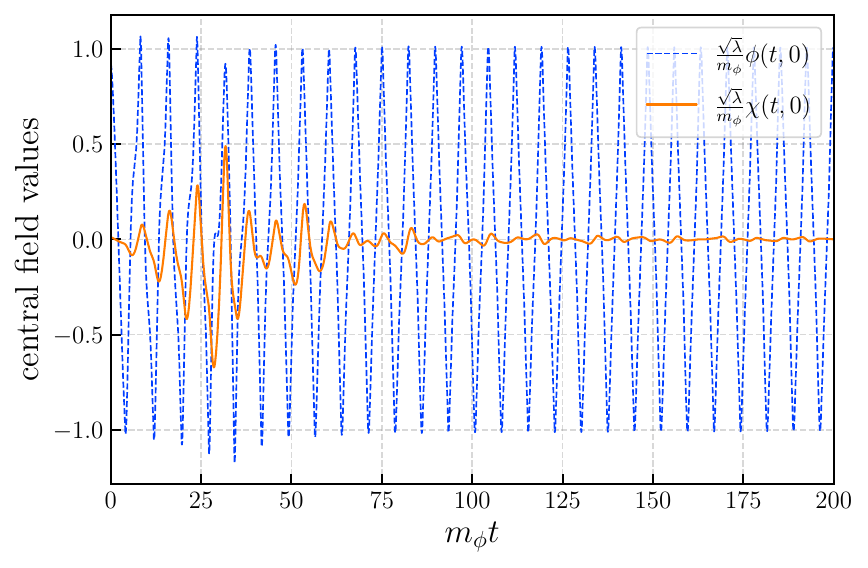}
    \end{subfigure}
    \hfill
    \begin{subfigure}{0.49\textwidth}
        \centering        \includegraphics[width=\textwidth]{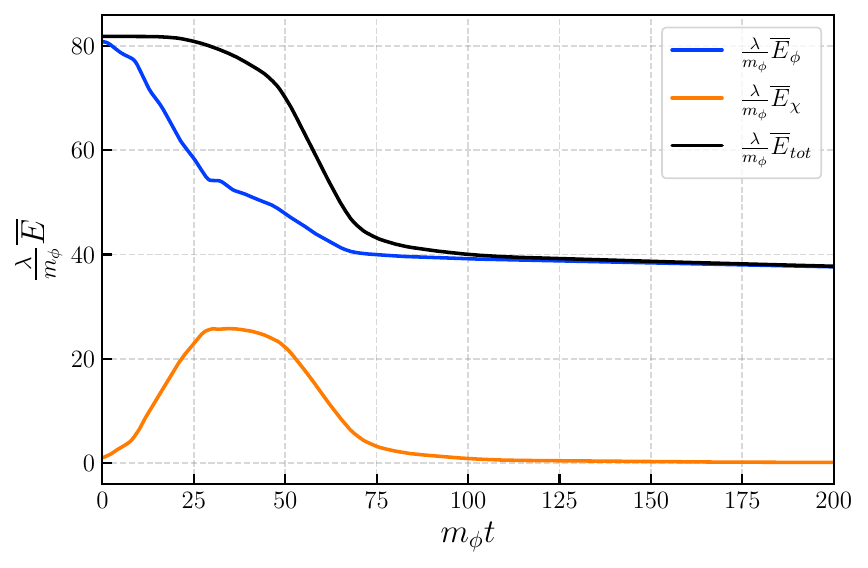}
    \end{subfigure}
    \caption{Time evolution of the centeral field values of $\phi$ and $\chi$, and the energies of both sectors in the two-field simulation.
    Top panel: initial oscillon frequency $\omega/m_\phi = 0.8$ with $m_\chi/m_\phi = 0.3$, $g_3/(m_\phi\sqrt{\lambda}) = 0.3$.
    Bottom panel: initial oscillon frequency $\omega/m_\phi = 0.8$ with $m_\chi/m_\phi = 0.6$, $g_3/(m_\phi\sqrt{\lambda}) = 0.7$.
    In both cases, $g_4 = 0$.}\label{fig:simulations g4=0}
\end{figure}

\begin{figure}[t]
    \begin{subfigure}{0.49\textwidth}
        \centering        \includegraphics[width=\textwidth]{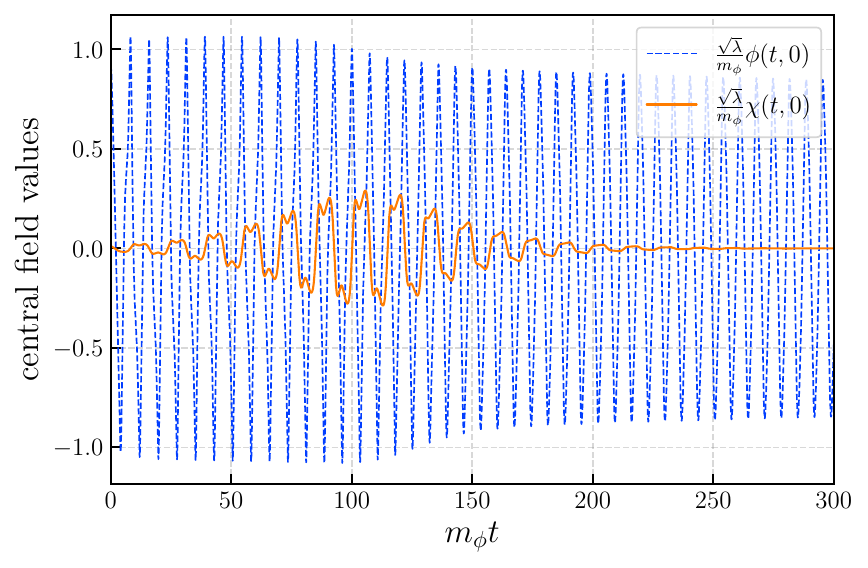}
    \end{subfigure}
    \hfill
    \begin{subfigure}{0.49\textwidth}
        \centering        \includegraphics[width=\textwidth]{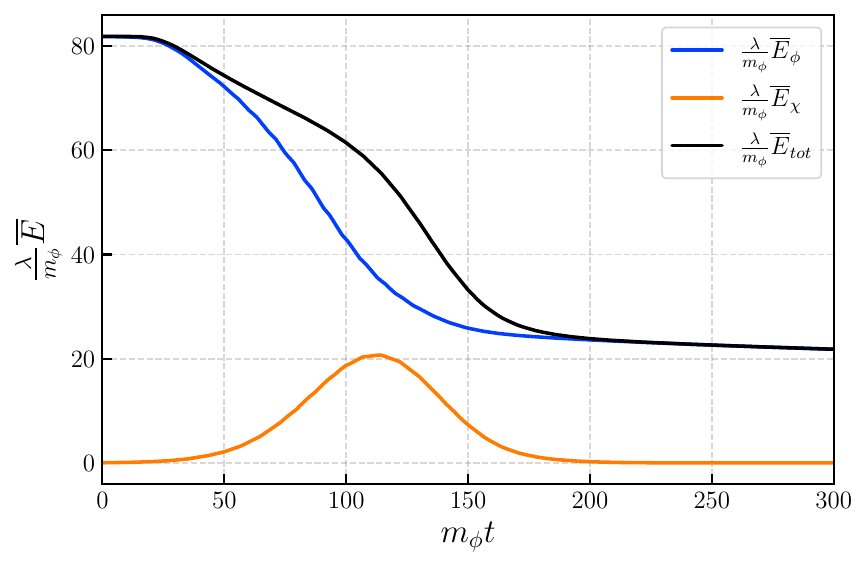}
    \end{subfigure}
    \caption{Time evolution of the centeral field values of $\phi$ and $\chi$, and the energies of both sectors in the two-field simulation, starting from an initial oscillon with frequency $\omega/m_\phi = 0.8$ , $m_\chi/m_\phi = 0.3$, $g_3/(m_\phi\sqrt{\lambda}) = 0.3$, and $g_4/\lambda=0.2$. 
    Compared with the top panel of Fig.~\ref{fig:simulations g4=0}, the presence of a nonzero $g_4$ modifies the instability of the $\chi$ field, so that the parametric resonance terminates before destroying the oscillon. }\label{fig:m3_34points}
\end{figure}

Figure~\ref{fig:simulations g4=0} shows the results of simulations for two parameter sets: (i) $m_\chi/m_\phi = 0.3$, $g_3/(m_\phi \sqrt{\lambda}) = 0.3$, starting from the initial profile of an oscillon with $\omega/m_\phi = 0.8$; and (ii) $m_\chi/m_\phi = 0.6$, $g_3/(m_\phi \sqrt{\lambda}) = 0.7$, starting from the initial profile of an oscillon with the same frequency.
In the top panel, the $\chi$ field undergoes rapid exponential growth due to parametric resonance, quickly extracting most of the energy from the $\phi$ oscillon.
In this case, since the critical energy $\overline{E}_{0}^{\textrm{osc}}$ is close to $E_{\textrm{death}}$, the oscillon is destroyed before the growth of $\chi$ ceases.
In contrast, in the bottom panel, the $\chi$ field exits the instability band after a short period of amplification as the configuration of $\phi$ evolves.

This behavior is consistent with our previous analysis of oscillons decay rate in Ref.~\cite{li_decay_2025}, which showed the fast decay into an external scalar field via parametric resonance is efficient only for large oscillons with $E>\overline{E}_{0}^{\textrm{osc}}$. 
The critical value $E>\overline{E}_{0}$ is determined by the coupling strength and mass of the daughter field.
When $\overline{E}_{0}^{\textrm{osc}} \gtrsim \overline{E}_{\textrm{death}}$, the dominant energy loss rate $\Gamma$ in each stage can be approximated as
\begin{align}
    \Gamma(\overline{E}) \sim 
    \begin{cases}
        \Re{(\tilde{\mu})} , \quad \overline{E} \gtrsim \overline{E}_{0}^{\textrm{osc}}\\
        \Gamma_{\xi} + \Gamma_{\textrm{per}}, \quad  \overline{E}_{\textrm{death}}< \overline{E} \lesssim \overline{E}_{0}^{\textrm{osc}},
    \end{cases}
\end{align}
where $\Gamma_{\textrm{per}}$ denotes the total perturbative decay rate of $\phi$ into $\chi$ through all interactions channels.
For the case  $\overline{E}_{0}^{\textrm{osc}} \lesssim \overline{E}_{\textrm{death}}$, we have
\begin{align}
    \Gamma(\overline{E}) \sim 
        \Re{(\tilde{\mu})} , \quad \overline{E} \gtrsim \overline{E}_{\textrm{death}}.
\end{align}

Figure~\ref{fig:m3_34points} presents simulation results for $m_\chi/m_\phi = 0.3$, $g_3/(m_\phi\sqrt{\lambda}) = 0.3$, and $g_4/\lambda=0.2$.
Compared with the top panel of Fig.~\ref{fig:simulations g4=0}, we find that a nonzero $g_4$ modifies the Floquet exponent even though the resonance is still primarily driven by the three-point interaction.
As a consequence,  the critical oscillon energy $\overline{E}_{0}^{\textrm{osc}}$ becomes smaller, so that the exponential growth of $\chi$ terminates before destroying the oscillon.
When both $g_3$ and $g_4$ are nonzero, the equation of motion for $\chi$ takes the form of a Hill-like equation with two oscillatory modes, each capable of inducing parametric resonance.
The resulting Floquet chart becomes more intricate in this case, as shown in App.~\ref{app:hills eq}.
The general method presented in the previous section can be applied to this case as well to determine the critical value of $\overline{E}_{0}^{\textrm{osc}}$ and its dependence on the oscillon profile and the coupling coefficients.
Since these quantities vary with the specific form of interaction considered, we do not repeat the detailed analysis here.
Nevertheless, the simulation shown in Fig.~\ref{fig:m3_34points} demonstrates that even in such more complex cases, there exist parameter regions where the parametric resonance of the external field ceases before depleting all the energy of the oscillon, which is valid for different forms of interactions.
This suggests that the termination of resonance prior to complete energy depletion is a generic and qualitatively interaction-independent feature.

\section{Discussions and conclusions}\label{sec:conclusion}

In this work, we addressed the question of how robust the oscillon decay dynamics through parametric resonance into an external scalar field, as identified in Ref.~\cite{li_decay_2025}, are to the specific form of the interactions. 
For this purpose, we have extended our previous investigation \cite{li_decay_2025} on oscillon decay and lifetime, which focused on a four-point coupling to an external scalar field~$\chi$, to include the case of a three-point interaction.
Our results provide further evidence that the decay of oscillons through parametric resonance, as well as the associated lifetime characteristics,are not tied to a specific form of coupling, but rather represent a general feature of oscillon-scalar-field interactions.

We analyzed the instability bands of the $\chi$ field and computed the corresponding Floquet exponent under an oscillating oscillon background.
Although the three-point coupling yields instability bands with shapes distinct from those in the four-point case, the qualitative dependence on the background oscillon configuration remains similar.
In particular, parametric resonance in the external field occurs only when the oscillon possesses sufficiently large energy and the coupling strength lies within a certain range.
The critical oscillon energy, $\overline{E}_{0}^{\textrm{osc}}$, depends on the coupling coefficients and the mass of the daughter field.

We also performed full two-field numerical simulations to examine the nonlinear dynamics.
As in the four-point case, we identified three distinct regimes depending on the initial oscillon energy $\overline{E}_{\textrm{ini}}$: (i) the oscillon collapses during the exponential amplification of the $\chi$ field when $\overline{E}_{\textrm{ini}} \gtrsim \overline{E}_{\textrm{death}} \gtrsim \overline{E}_{0}^{\textrm{osc}}$; (ii) for $\overline{E}_{\textrm{ini}} \gtrsim \overline{E}_{0}^{\textrm{osc}} \gtrsim \overline{E}_{\textrm{death}}$, the resonance of the $\chi$ field terminates before the oscillon is destroyed, leaving a residual oscillon that decays perturbatively until its eventual ``energetic death''; (iii) no significant parametric resonance occurs if $ \overline{E}_{0}^{\textrm{osc}} \gtrsim \overline{E}_{\textrm{ini}} \gtrsim \overline{E}_{\textrm{death}} $.
We further confirmed that these behaviors persist when both three- and four-point couplings are present, although the detailed structure of the instability bands and parameter dependence become more intricate.

Overall, our findings imply that while the exact critical oscillon energies depend on the specific form of the coupling, the qualitative behavior is universal: decay through parametric resonance of an external scalar field does not necessarily destroy the oscillon but instead extracts a finite portion of its energy within an appropriate coupling range.
This robustness across interaction types suggests that the persistence of oscillons after inflation play a nontrivial role in the reheating process, potentially leading to richer and more spatially inhomogeneous post-inflationary dynamics.

\section*{Acknowledgments}
We are grateful to Masahide Yamaguchi for valuable discussions and for kindly reviewing the draft of this paper.
S.L. is supported by JSPS Grant-in-Aid for Research Fellows Grant No.23KJ0936, and by IBS under the project code, IBS-R018-D3.

\appendix

\section{Hill's equation with \texorpdfstring{$g_3 \neq 0,\ g_4 \neq 0$}{g3 != 0, g4 != 0}} \label{app:hills eq}

\begin{figure}[t]
    \centering        \includegraphics[width=0.6\textwidth]{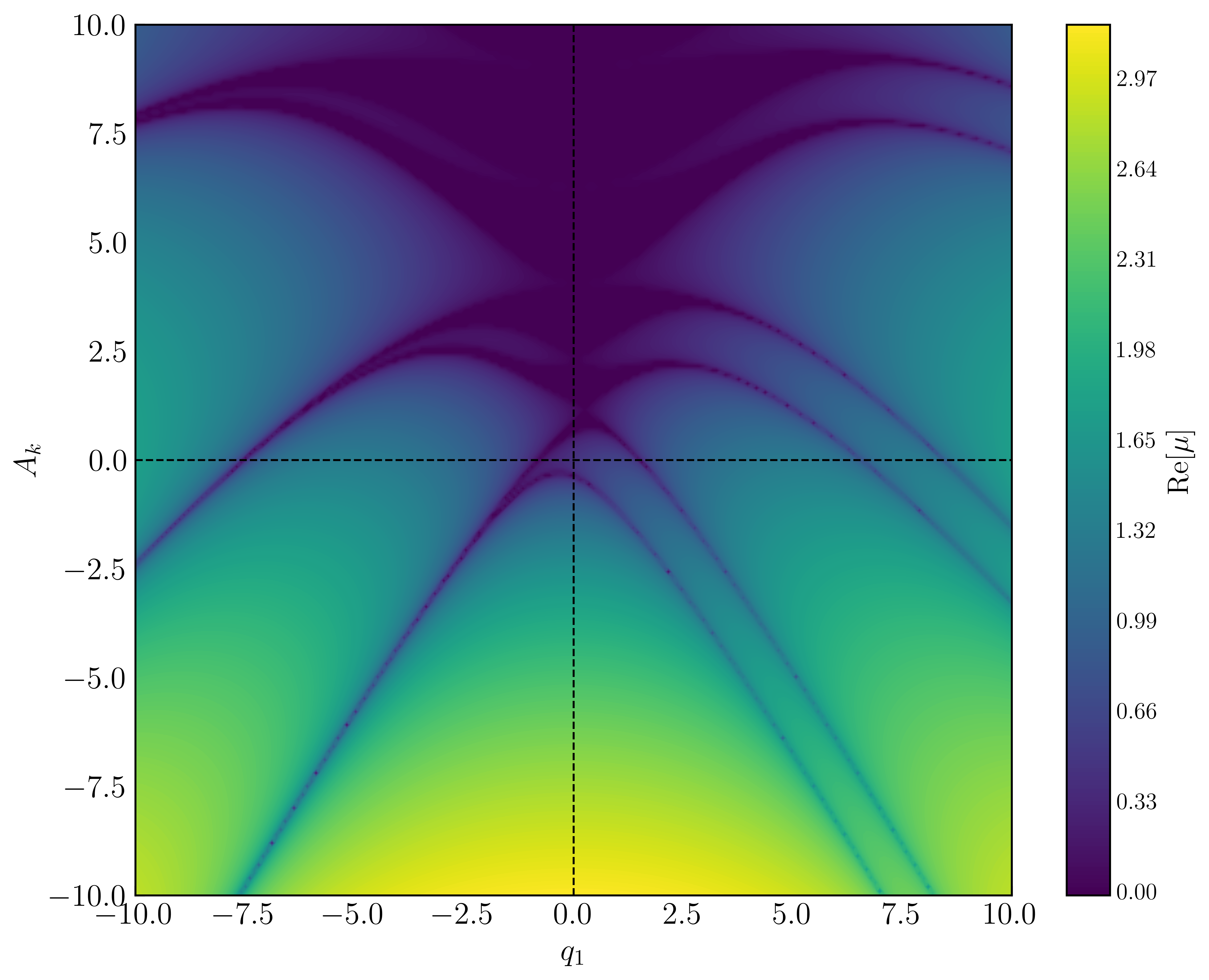}
    \caption{Floquet chart for the Hills' equation in Eq.~\eqref{eq:Hill's eq} shown in terms of $A_k$ and $q_1$ with $q_2=0.5$.}\label{fig:hill's eq}
\end{figure}

When $g_3 \neq 0,\ g_4 \neq 0$, the equation of $\chi_k$ under the homogeneous approximation can be organized to a general form of Hill's equation, where more than one driving modes are responsible for the parametric resonance,
\begin{equation}
\begin{aligned}
    &\chi_k^{''} + (A_k + 2 q_1 \cos{(z)} + 2 q_2 \cos{(2z)})\chi_k  = 0,\\
    A_k &= \frac{k^2 + m_\chi^2}{\omega^2} + 2q_2,~~ q_1 = \frac{2g_3\psi_0}{\omega^2},~~q_2 = \frac{2g_4\psi_0^2}{\omega^2}.~~ z = \omega t,\label{eq:Hill's eq}
\end{aligned}
\end{equation}
In this case, the Floquet chart becomes more intricate.
As an example shown in Figure~\ref{fig:hill's eq} in terms of $A_k$ and $q_1$ with $q_2 = 0.5$, additional stability tongues appear, dividing the instability bands into more pieces.
These structures lead to more complex dependence on the physical quantities we are interested in.
Therefore, specific analysis for the parametric dependence is necessary when considering different forms of interaction.

\bibliographystyle{JHEP}
\bibliography{three-point}

\end{document}